# Electronical Health Record's Systems. Interoperability

**Univ.Assist. Apostol Angela Simona,**
**"Tibiscus"University of Timişoara, România**
**Physicist Câtu Cosmin,**
**University of Medicine and Pharmacy, Timişoara, România**
**Assist.Prof. Vernic V. Corina, Ph.D.**
**University of Medicine and Pharmacy, Timişoara, România**

ABSTRACT: Understanding the importance that the electronic medical health records system has, with its various structural types and grades, has led to the elaboration of a series of standards and quality control methods, meant to control its functioning. In time, the electronic health records system has evolved along with the medical data's change of structure. Romania has not yet managed to fully clarify this concept, various definitions still being encountered, such as "Patient's electronic chart", "Electronic health file". A slow change from functional interoperability (OSI level 6) to semantic interoperability (level 7) is being aimed at the moment. This current article will try to present the main electronic files models, from a functional interoperability system's possibility to be created perspective.

## 1. Introduction

Modern medical systems are nowadays based more and more on the easy access and real time processing of a great amount of data and medical information characterized by certain grade and type of structuring. Understanding the major importance that electronic data base systems have for the medical informatics systems led to their quick worldwide adoption, not to mention that this particular process is being sustained by certain standards and quality management methods. One can not afford to overlook the difficulties and the fear associated with the migration of electronic medical database among which: technical limitations in what security is concerned, integrity and access to data, high costs, the lack of operating skills and trust in computerized systems along with the opposition to the new change, small diversity of software products, the absence of certain quality standards generally accepted and the lack of adequate laws.

In time, the concept of electronic database evidence has suffered many transformations and reconsiderations along with the increase in complexity and with the change of medical data, simultaneous with health care systems evolution. Romania has not yet managed to fully understand this concept, therefore we can encounter definitions such as "patient's electronic chart", "electronic health file". Since there is no consent for that matter, any reference to electronic health records system will be made according to the ISO/TR 20514 definition, using the abbreviation EHR: "Collection of computerized information referring to the health state of a certain subject stored and transmitted in complete safety, accessible to any authorized user. It has a logic pattern for information organization implemented, universally accepted and independent from the system. Its main aim is to assure continuous, efficiently and quality integrated health services along with retrospective and prospective information ".

The terminological variation, often implying national specific, not only depends on the formulation but also on the content and clinical context. The electronic health file content still brings true debates to life. Obviously, the content will depend on the aim, type of requested information, amount of time, access rights and last but not least on the balance between the included information's quantity and their costs. A complete EHR system must include demographic data, notes on the health state of the patient, illness, medication, vital signs, history of the disease as well as history of the family diseases, laboratory data, immunization, and imagistic data.

A high quality EHR system must be perfectly adapted to the context and it consists of several interoperable information organizing levels. The implementation details vary a lot according to the level of medical assistance (primary assistance, ambulatory assistance, emergency, hospital etc), to the standards that are respected, to the infrastructure that is based on as well as some other factors.

Those generically qualities stand for the definition of a more detailed functional and structural request of an EHR. The attendance of functional requests as well as the security of an EHR is legislated in some states such as USA (where functions a single organism of certifying - CCHIT), Canada,

Belgium, Ireland, Denmark and England. Organizations such as Continua Health Alliance and the initiative of IHE (Integrating the Healthcare Enterprise) - supported by HIMMS (Healthcare Information and Management Systems Society) are oriented towards the auto evaluation of technical support of the products by the consumer. In Europe, with the help of the European Committee, the Eurorec Institute fights for developing a European mechanism for evaluation and certification of the EHR systems, having the functional criteria and security matters in mind. Of great importance for that matter are the research projects financed by Frame Program 6 developed by Eurorec, especially the QRec project (European Quality Labelling and Certification of Electronic Health Record systems) and EHR Implement. All those can be defined according to the management of quality, normative evaluation emphasizing on monitoring and quantification of quality, putting beside the evaluation research methods based mainly on methodological and theoretical expertise.

Nowadays, the efforts are focused on gradual passing from functional interoperability (level 6 OSI) towards semantically interoperability (level 7). The easy access and sort out of medical data must be completed by incorporating efficient security mechanisms in the EHR systems, meant to ensure integrity and confidentiality of the stored data.

## 2. The electronic health file

The electronic health file assures the user to access the patient's health related information from a certain database. These patients data base will periodically be updated based on the information received from various medical departments. The file can only be accessed by the medical staff and personnel that the patient agrees with.

The electronic health file aims to be able to give online medical information related to certain patients. The information given to the user are accessible according to several security levels, each user only having access to the information that he has the right for.

The patient has access to several details, of course, technical details that he does not understand not being included. One of the first conditions for accomplishing this system was the patient's ability to see who has accessed his data as well as who had the endeavor to access the data base in order to ensure more security means. Recent national systems exist, but when it comes to complete spread, no country has succeeded this performance. Among the countries that already have functional projects, the European model that was the ground for the pilot project in Romania must be remembered, the English model, extremely well adjusted and the American one, completed mostly in the same period with the European one.

## 3. The electronically evidence of medical data – terminology

An electronic health record or as it is known in Romania, electronic health file, refers to all the records related to the health state of a certain individual along his life in a digital format and it is the equivalent for the traditional file. The purpose is to track the coordination of storing systems so that the finding of the electronic health record could be easy accessed with the help of computers. Electronic health records are usually accessed from a computer, mainly through a network. It can consist of electronic medical recordings from different locations or sources. A variety of information linked to medical assistance services can be stored and accessed in this manner.

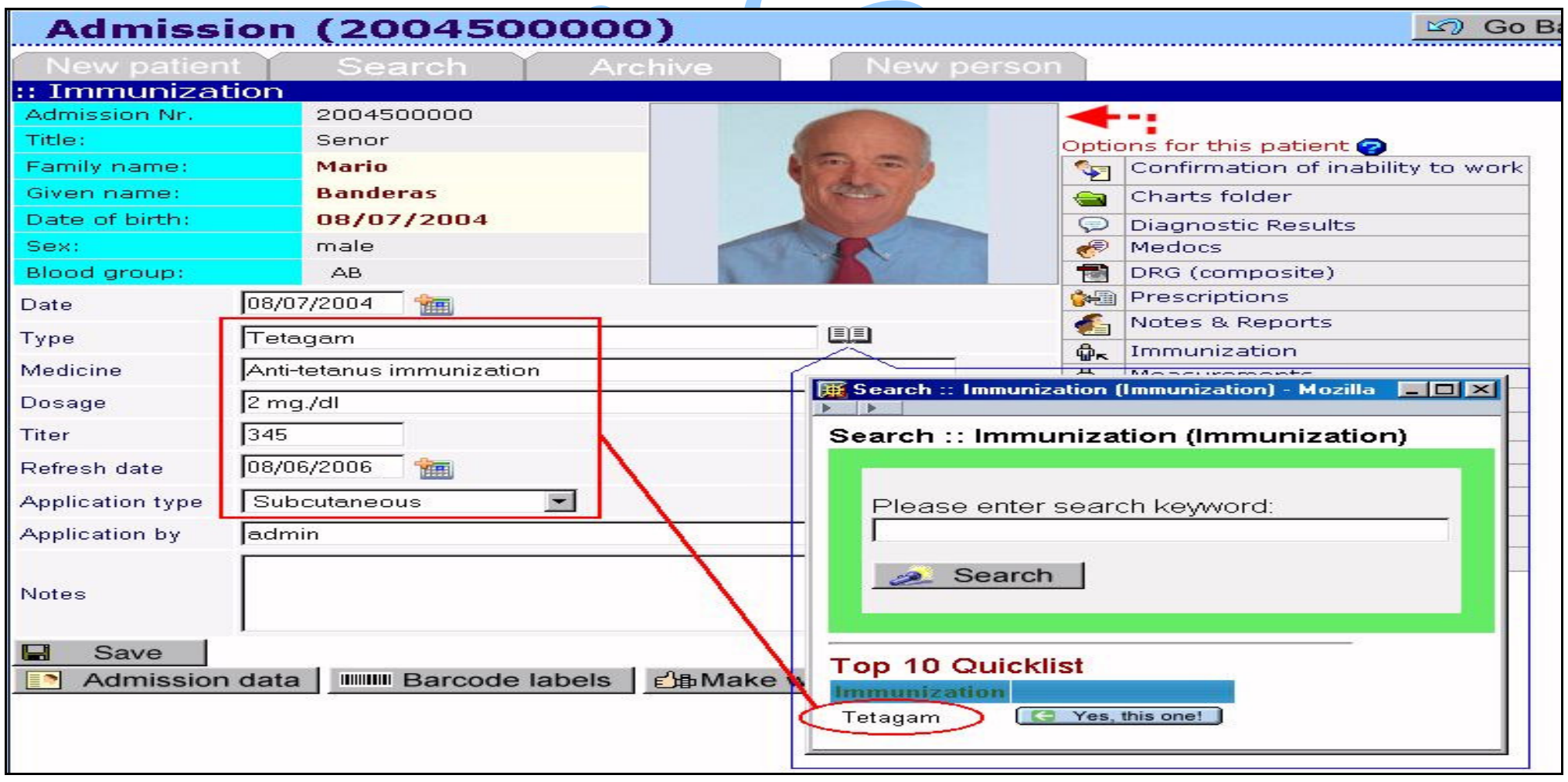

***Figure 1. Electronic health record (EHR)***

The DES systems can reduce medical errors. It was believed that DES systems will increase the efficiency and lower costs, as well as promote the standardization of medical care services. Even though EMR , providers of computerized systems, have existed for more than 30 years in this business, less than 10% of the hospitals in 2006 had a complete system.

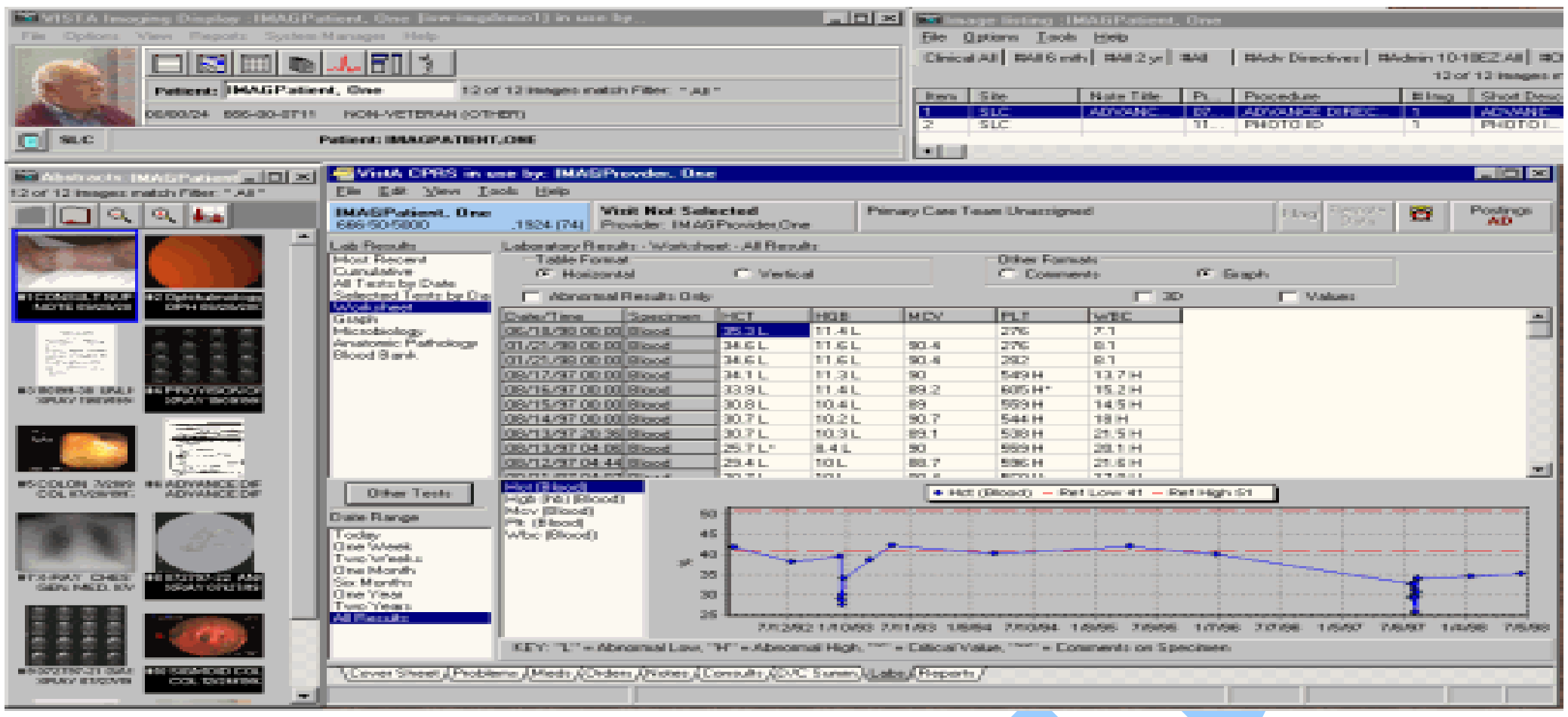

***Figure 2. Sample patient record view from an image-based electronic health record (VistA)***

A multitude of systems have been used in order to define the electronic health record of the patient, in the end leading to a superposition of definitions. Both terms, electronic health file as well as electronic medical record are being used. Some of the medical software users correlate the electronical medical record term to a global level, and the electronic medical record term to a recording. For most users, both terms are being used in the same simultaneously.

In time, the electronic evidence of medical data concept has suffered several changes along with the increase in complexity and natural novelties, and with the evolution of health care systems.

The variations in terminology, often implying the national characteristic, are not only connected to formulation but also to the content and clinical context. Nowadays this concept has many forms.

***EHR/EHCR (electronic health-care- record)*** - Europe – represents a more and more accepted concept which extends the signification of the electronic health care record beyond the illness of the patient, to an entire health history along with the actual state of the patient and the future supervision plans.

The change of medical data and the evolution that the health care systems have suffered led to a number of changes in what the concept of electronic data base evidence is concerned. “Patient’s electronic chart”, “electronic health file” are definitions that can still be encountered in Romania, since the country has not yet managed to fully clarify the concept. According to the ISO/TR definition, any reference to electronic health records will be made using the EHR abbreviation.

***PCR (patient-carried record)*** – experimented model in various countries in the form of medical cards, so that all the medical information belong to the beholder. The subject of the record is the patient; a specialization on clinical domains for the recorded data however exists. The current trend is for these cards to contain a minimum of necessary information, in order for the user to be recognized (patient as well as medical staff), all other data regarding the patient being stored in a distributed data base.

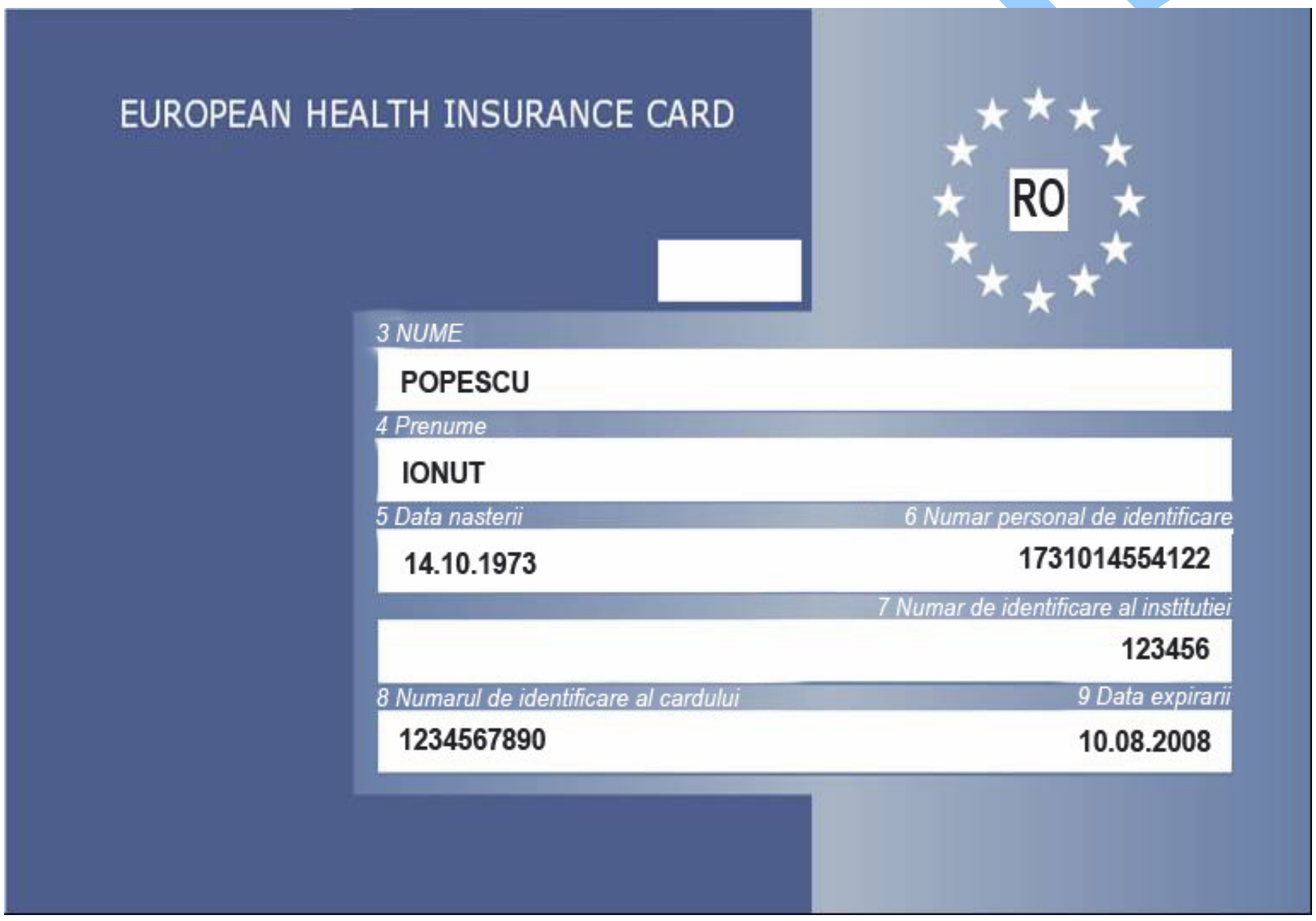

***Figure 3. European Card of Health Social Assurance***

***CMR (computerized medical record)***- a general concept referring to any form of electronic data base evidence starting from simple Word documents or Excel and up to various technical administration of the medical data implementations, more or less complex.

***CPR (computer-based patient record)*** – USA – it ensures the patient's entire life medical data recording. It implies medical data acquisition from different sources (family doctor, dentist, specialists) which requires a high grade of data interoperability, hard to achieve at the moment. It must be noticed that the records refer to a single subject- the patient.

EPR (electronic patient record) – England- is similar to CPR with the only difference that it is focused on medical data belonging to one or more specialties (general medicine, dental medicine, surgery, intensive care unit, etc. and without the desire to cover the patients whole life. EPR can be considered as a subset (at a physical or logical level) of the EHR.

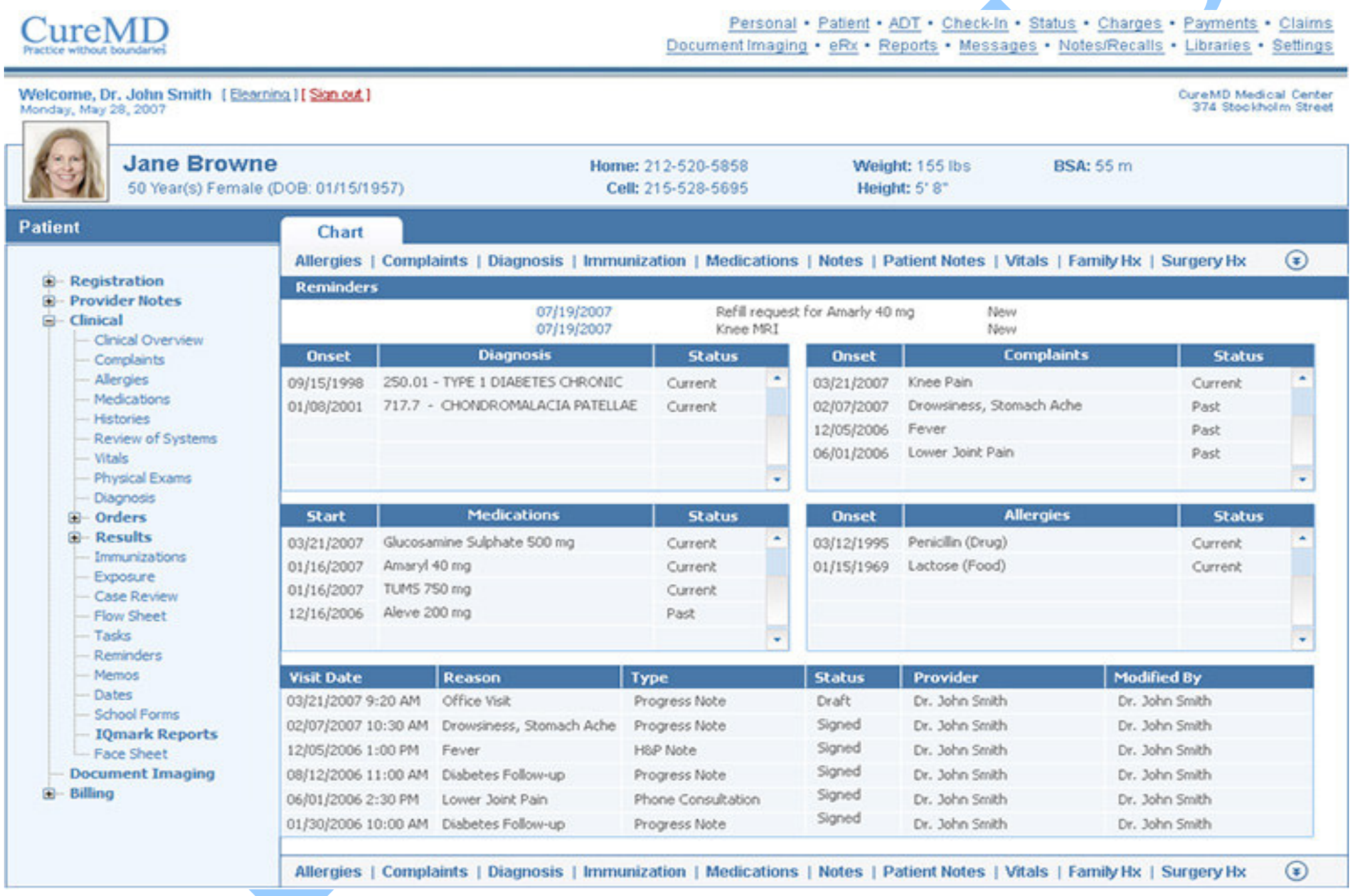

***Figure 4. Computerized medical record***

***EMR (electronic medical record)*** – New Zeeland- consists of a record card for medical data based on a total interoperability in a medical institution (hospital, dispensary). The subject of the recording is the patient, and his file has a classical approach.

***PHR (personal health record)*** – consists of an electronic file initiating a physical person. An ideal PHR should offer a complete summary and an accurate view on the state of health of the patient together with an individual medical history by collecting data from different sources. A PHR should make this information online available for everyone with electronic accreditation to see the information.

EHR systems are able to reduce medical errors [1]. In one ambulatory healthcare study, there was no difference in 14 measures, improvement in 2 outcome measures, and worse outcome on 1 measure [2].

EHR systems are supposed to increase physician efficiency and reduce costs, as well as promote standardization of care.

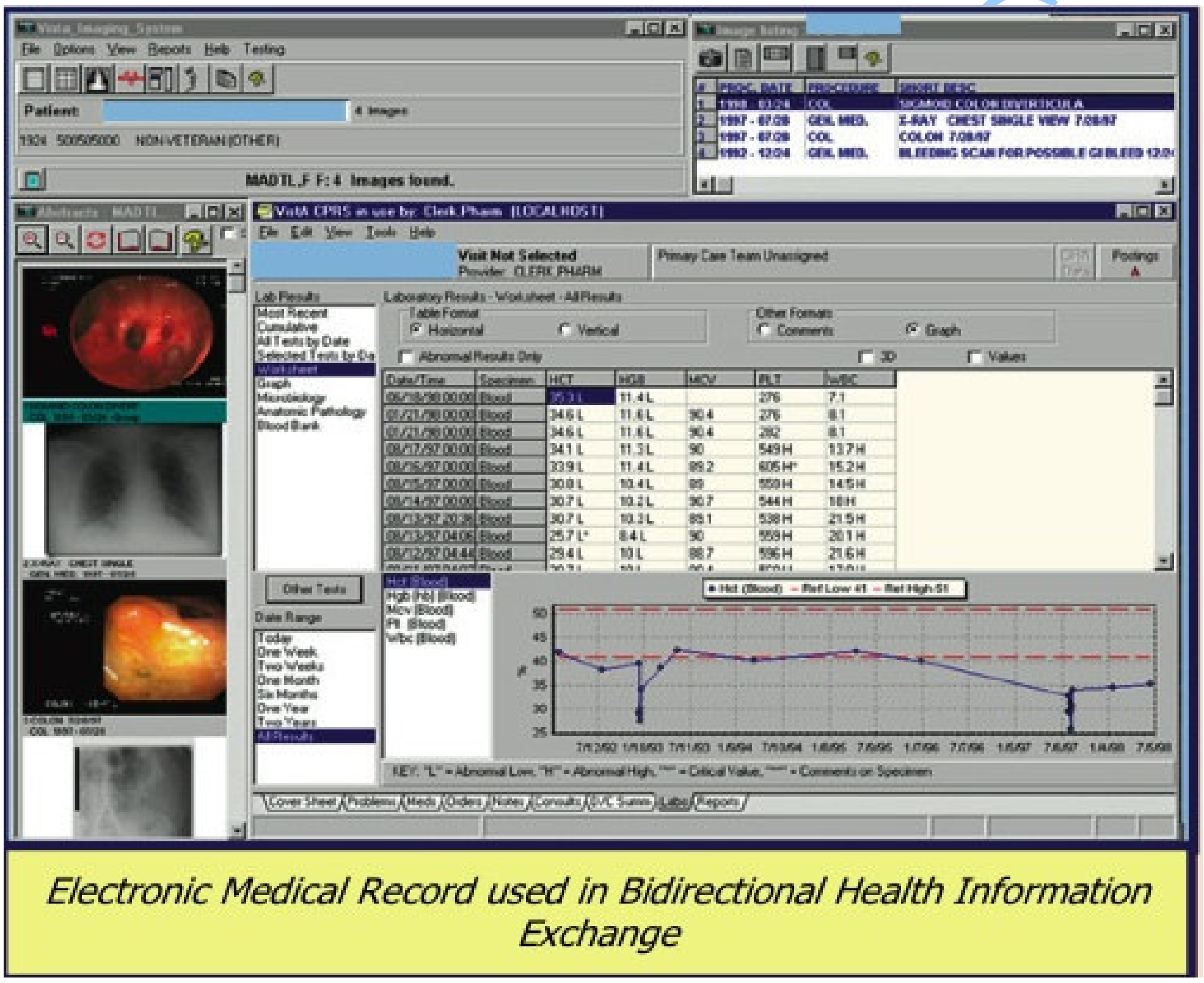

***Figure 5. Electronic Medical Record***

## 4. Types of data stored in an electronic medical record

An electronic medical record might include:

- Patient demographics.

- Medical history, examination and progress reports of health and illnesses.
- Medicine and allergy lists, and immunization status.
- Laboratory test results.
- Radiology images (X-rays, CTs, MRIs, etc.)
- Photographs, from endoscopy or laparoscopy or clinical photographs.
- Medication information, including side-effects and interactions.
- Evidence-based recommendations for specific medical conditions
- A record of appointments and other reminders.
- Billing records.
- Eligibility
- Advanced directives, living wills, and health powers of attorney

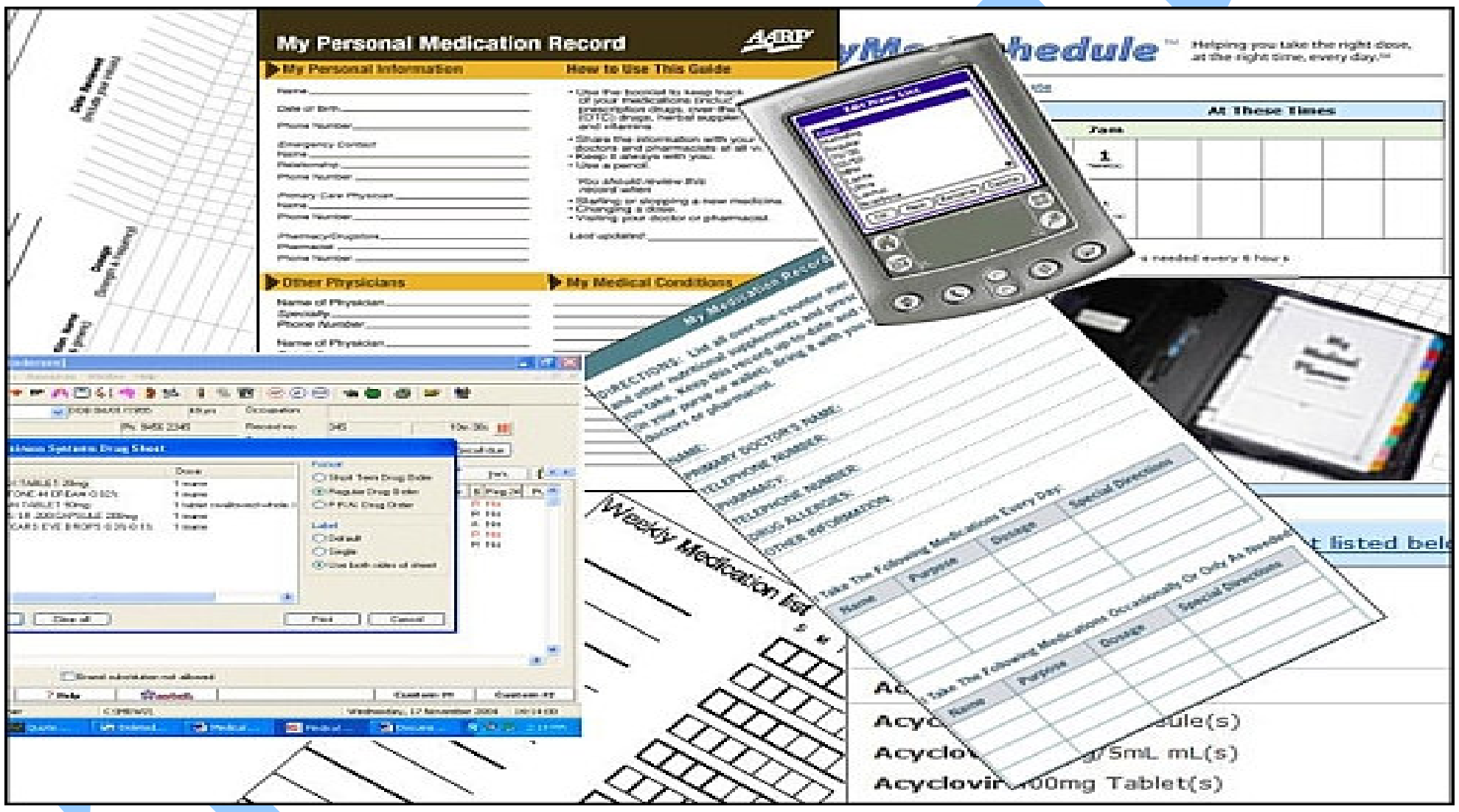

*Figure 6. Personal Health Record*

## 5. Ideal characteristics of an electronic health record (EHR)

- Information should be able to be continuously updated.
- The data existing in an electronic health record system should be able to be used anonymously for statistical reporting, for purposes of quality improvement, outcome reporting, resource management, and public health communicable disease surveillance [3].

- The ability to exchange records between different electronic health records systems ("interoperability" [4]) would help develop a network between healthcare deliveries in non-affiliated healthcare facilities.

## 6. Interoperability

In healthcare, interoperability refers to the ability of different information, technology systems and software applications to communicate, to exchange data between them, accurately, effectively, and especially to use the information that has been exchanged.

## 7. Organizations to evaluate standardization proposals

Several models of standardization for electronic medical records and electronic medical record exchange have been proposed and multiple organizations formed to help evaluate and implement them [5][6].

## 8. Organizations

- CHI (Consolidated Health Informatics Initiative) - recommends nationwide federal adoption of EHR standards in the United States
- CCHIT (Certification Commission for Healthcare Information Technology) - a federally funded, not-for-profit organization that evaluates and develops the certification for EHRs and interoperable EHR networks (USA)
- IHE(Integrating the Healthcare Enterprise) - a consortium, sponsored by the HIMSS, that recommends integration of EHR data communicated using the HL7 and DICOM protocols
- ANSI(American National Standards Institute) - accredits standards in the United States and co-ordinates US standards with international standards
- Healthcare Information and Management Systems Society(HIMSS) - an international trade organization of health informatics technology providers
- American Society for Testing and Materials - a consortium of scientists and engineers that recommends international standards

- openEHR - provides open specifications and tools for the 'shared' EHR
- Canada Health Infoway - a federally funded, not-for-profit organization that promotes the development and adoption of EHRs in Canada
- World Wide Web Consortium(W3C) - promotes Internet-wide communications standards to prevent market fragmentation
- Clinical Data Interchange Standards Consortium(CDISC) - a non-profit organization that develops platform-independent healthcare data standards
- EHR-Lab Interoperability and Connectivity Standards (ELINCS) - run by the HL7 group to help provide lab data and other EHR Interoperability

## 9. Standards

- ANSI X12 (EDI) - transaction protocols used for transmitting patient data. Popular in the United States for transmission of billing data.
- CEN's TC/251 provides EHR standards in Europe including:
  - EN 13606, communication standards for EHR information
  - CONTSYS (EN 13940), supports continuity of care record standardization.
  - HISA (EN 12967), a services standard for inter-system communication in a clinical information environment.
- Continuity of Care Record- ASTM International Continuity of Care Record standard
- DICOM - an international communications protocol standard for representing and transmitting radiology (and other) image-based data, sponsored by NEMA(National Electrical Manufacturers Association)
- HL7 - a standardized messaging and text communications protocol between hospital and physician record systems, and between practice management systems
- ISO - ISO TC 215 provides international technical specifications for EHRs. ISO 18308 describes EHR architectures

## 10. Synchronization of records

In the case of providing care in two different facilities, it may be difficult to update records at both locations. This is a problem that plagues distributed computer records in all industries.

There are two models conceived in order to satisfy this problem: a centralized data server solution and a peer-to-peer file synchronization program (as has been developed for other peer-to-peer networks).

In the United States, Great Britain, and Germany, the concept of a national centralized server model of healthcare data has not been received successfully yet. There are still several issues of privacy and security that need to be solved [7][8].

Synchronization programs for distributed storage models, however, are only useful once record standardization has occurred. It is a challenge to try to merge already existing public healthcare databases. The key benefit consists in the ability of electronic health record systems to provide this function and will help improve healthcare delivery [9][10][11].

## 11. Successful implementations of EHR systems

In the United States, the Department of Veterans Affairs (VA) has the largest enterprise-wide health information system that includes an electronic medical record, known as the Veterans Health Information Systems and Technology Architecture or VistA. A graphical user interface known as the Computerized Patient Record System (CPRS) allows health care providers to review and update a patient's electronic medical record at any of the VA's over 1,000 healthcare facilities. CPRS includes the ability to place orders, including medications, special procedures, X-rays, patient care nursing orders, diets, and laboratory tests.

The US Indian Health Service uses an EHR similar to VistA called RPMS. VistA Imaging is also being used to integrate images and co-ordinate PACS into the EHR system.

As of 2005, the National Health Service (NHS) in the United Kingdom also began an EHR system. The goal of the NHS is to have 60,000,000 patients with a centralized electronic health record by 2010. The plan involves a gradual roll-out commencing May 2006, providing general practitioners in England access to the National Programme for IT (NPfIT) [12].

The Canadian province of Alberta started a large-scale operational EHR system project in 2005 called Alberta Netcare, which is expected to encompass all of Alberta by 2008.

## 12. Software criteria of interoperability

The Center for Information Technology Leadership described four different categories of data structuring at which health care data exchange can take place.[13] Each has different technical requirements and offers different potential for benefits realization.

The four levels are [14]:

| Level | Data Type | Example |
|---|---|---|
| 1 | Non-electronic data | Paper, mail, and phone call. |
| 2 | Machine transportable data | Fax, email, and unindexed documents. |
| 3 | Machine organizable data (structured messages, unstructured content) | HL7 messages and indexed (labeled) documents, images, and objects. |
| 4 | Machine interpretable data (structured messages, standardized content) | Automated transfer from an external lab of coded results into a provider's EHR. Data can be transmitted (or accessed without transmission) by HIT systems without need for further semantic interpretation or translation. |

## References

**[1].** "The Best Medical Care In The U.S. How Veterans Affairs transformed itself – and what it means for the rest of us" (July 16, 2006). *BusinessWeek, Red Oak, IA (USA).*

**[2].**Linder JA, Ma J, Bates DW, Middleton B, Stafford RS (Jul 2007). "Electronic health record use and the quality of ambulatory care in the United States". *Arch. Intern. Med.* **167** (13): 1400–5. doi:10.1001/archinte.167.13.1400. PMID 17620534.

**[3].** Healthcare Information and Management Systems Society (2003): {{PDF|EHR Definition, Attributes and Essential Requirements|152 KiB Retrieved July 28, 2006

**[4].** Adapted from the IEEE definition of interoperability, and legal definitions used by the FCC (47 CFR 51.3), in statutes regarding copyright protection (17 USC 1201), and e-government services (44 USC 3601)

**[5].** Nainil C. Chheda, MS (January 2007). "Standardization & Certification: The truth just sounds different" (PDF). *Application of Healthcare Governance*. Retrieved on 2007-01-16.

**[6].** Nainil C. Chheda, MS (November 2005). "Electronic Medical Records and Continuity of Care Records - The Utility Theory" (PDF). *Application of Information Technology and Economics*. Retrieved on 2006-07-25.

**[7].** "Opposition calls for rethink on data storage". e-Health Insider (UK) (December 2007).

**[8].** "German doctors say no to centrally stored patient records". e-Health Insider (UK) (January 2008).

**[9].** "Integrating the New York citywide immunization registry and the childhood blood lead registry" (2004). *Journal of Public Health Management and Practice*: S72–80.The Master Child Index consolidated 4,610,585 records that were contained in both databases into 2,977,290 records through a match and merge system.

**[10].** "Quality improvements in pediatric well care with an electronic record" (2001). *Proc AMIA Symp*: 209–13.

**[11].**"Perspectives on integrated child health information systems: Parents, providers, and public health" (2004). *Journal of Public Health Management Practice*: S57–S60.

**[12].** NHS Connecting for Health:Delivering the National Programme for IT Retrieved August 4, 2006

**[13].**Walker J, et al. (2005-01-19). "The Value Of Health Care Information Exchange And Interoperability". *Health Affairs*. doi:10.1377/hlthaff.w5.10. PMID 15659453.

**[14].** NAHIT Levels of EHR Interoperbility. "What is interoperability?". National Alliance for Health Information Technology. Retrieved on 2007-04-04.